# Reconciling the Lack of a Robust Anvil Cloud Amount Response to Warming

Zachary McGraw[1], Blaž Gasparini[1], Wouter Mol[1], Aiko Voigt[1*]

*1 Department of Meteorology and Geophysics, University of Vienna, Austria*

** Now at Department of Meteorology, University of Bonn, Germany*

*Corresponding author: Zachary McGraw, zachary.mcgraw@univie.ac.at*

**Key Points**

- Anvils experience more stability under warming because they rise into thinner air, which does not weaken the detrainment supplying them
- Convectively driven motion stays nearly fixed along isotherms, so warming relocates detrainment upward with little reduction
- Across RCEMIP-I models, anvil-level stability robustly increases, while anvil amount and detrainment of air and water respond far less

**Abstract |** The stability-iris hypothesis proposes that tropical anvil cloud amount decreases with warming as anvils rise into environments of greater static stability. Here we show that there is no substantial theoretical or simulated link from increased static stability to reduced anvil amount. We demonstrate analytically that the sensitivities of static stability and pressure velocity to surface warming arise mainly from decreasing air density, which does not weaken the detrainment supplying anvils. Convectively driven motion stays nearly fixed along isotherms, so detrainment changes little where anvils rise. An ensemble of RCEMIP simulations confirms that stability increases robustly but is not accompanied by a robust reduction in the detrainment of air or water at anvil levels. Overall, these findings dispel the expectation of a systematic reduction in anvil amount with warming arising from basic thermodynamic arguments.

**Plain Language Summary |** Tropical anvil clouds are important for regulating Earth's climate, but it remains uncertain how they will change as the planet warms. A leading hypothesis proposes that warming makes the atmosphere more stable where anvil clouds form, suppressing the upward motion that supplies these clouds and causing their amount to decrease. We test this idea using basic physical relationships and simulations from several convection-resolving models. We find that the increase in atmospheric stability as anvil clouds rise with warming is largely a consequence of reduced air density at higher altitudes, and is not a change that weakens the upward motions or water supply that form anvil clouds. Consistent with this finding, the simulations show little change in anvil amount or the supply of air or water to anvil levels. Hence, increased stability gives no clear reason to expect tropical anvil clouds to systematically decrease as Earth warms.

## 1. Introduction

Tropical anvil clouds are regarded as a key influence on Earth's temperature sensitivity, owing to their extensive coverage, strong radiative influence, and uncertain response to warming (Sherwood et al., 2020; Forster et al., 2021). It is established that anvil clouds amplify warming by rising to altitudes at which they more effectively trap outgoing radiation (Hartmann and Larson, 2002; Zelinka and Hartmann, 2011; Gasparini et al., 2024; Voigt et al., 2024; McGraw et al., 2025). Yet the potential for climate feedbacks due to anvil amount responses has been debated for decades (Lindzen et al., 2001; Lin et al., 2002; Lindzen and Choi, 2022; McKim et al., 2024).

The most widely accepted hypothesis for an altered anvil cloud amount is the stability-iris mechanism of Bony et al. (2016), based on earlier work by Zelinka and Hartmann (2010, 2011). The hypothesis proposes that warming decreases anvil amount by raising the anvils into layers of stronger static stability. This framework combines the approximate isothermal rise of anvils (Hartmann and Larson, 2002) with the constraint that, in subsiding regions, clear-sky atmospheric radiative cooling ($Q_r$, in K/s, defined here positive for cooling) is roughly balanced by adiabatic subsidence warming, such that

$$\omega \approx Q_r / S \qquad \text{Eqn. 1}$$

where $\omega$ (in Pa/s) is the velocity of subsiding air in pressure coordinates and $S$ (in K/Pa) is the pressure-coordinate static stability. This follows from the weak temperature gradient approximation (Sobel and Bretherton, 2000), whereby horizontal temperature gradients in the tropics are small enough to neglect in the thermodynamic equation. Bony et al. (2016) then define a radiatively driven divergence ($D_r$, in day-1) as the vertical divergence of this subsidence,

$$D_r = \frac{\partial (Q_r / S)}{\partial p} \qquad \text{Eqn. 2}$$

which is positive in the upper troposphere, where subsidence weakens with height. There, mass continuity requires subsiding regions to draw in air horizontally from convective regions. $D_r$ thus provides a radiative estimate of convective detrainment: the rate at which air mass leaves convection, per unit mass of air. From this, Bony et al. propose that $D_r$ controls anvil cloud amount, with increased stability thereby reducing anvil cloud amount under warming. Using a cloud ice budget, Beydoun et al. (2021) physically linked detrainment to anvil area and attributed a modeled anvil reduction to declining $D_r$, while further support has been identified in observations and models (Cronin & Wing, 2017; Saint-Lu et al., 2020, 2022). Other studies, however, have found the simulated anvil response to vary in sign (Ohno & Satoh, 2018; Wing et al., 2020) or to be weak (McKim et al., 2024; Deutloff et al., 2026).

This lack of consensus raises a basic question: *is there truly a fundamental stability-modulated anvil cloud amount sensitivity to surface warming?* Here we revisit the original theoretical arguments of the stability-iris hypothesis and examine radiative-convective simulations. We show that the robust increase in stability mainly reflects the lower density of air as anvils rise in response to surface warming, which we demonstrate cannot be expected to result in substantially reduced detrainment. The basis for the stability-iris hypothesis therefore does not hold.

## 2. Methods

To test the stability-iris hypothesis, we analyze simulations from Phase I of the Radiative-Convective Equilibrium Model Intercomparison Project (RCEMIP-I; Wing et al., 2018a, 2020). Their idealized configuration and explicitly resolved convection make them well suited for examining the physical basis of the hypothesis, following earlier RCE studies of the mechanism (Cronin & Wing, 2017; Wing et al., 2020; Jeevanjee, 2022). To quantify sensitivities to warming, we analyze the simulations with uniform sea surface temperatures ($T_s$) of 295 and 305 K.

To analyze detrainment and its sensitivity to warming, we separate air and water fluxes into convective and subsiding regions. Following Williams and Jeevanjee (2025), convective cells are grid points with cloud condensate above $10^{-5}$ kg/kg and upward velocity above 1 m/s, in 3D snapshots taken every 6 h over the final 25 days of each simulation. The subsiding region comprises all non-convective cells, with its mean subsidence velocity $w$ (m/s) defined as the convective updraft velocity summed over area and divided by the domain area. This upward volume flux by convection (McGraw & Gasparini, 2026) closely matches the mean descent of non-convective cells, which occupy at least 99 % of grid cells at any level. Air detrainment is the decrease of $w$ with height, $-\partial w/\partial z$, or in pressure coordinates $\partial \omega/\partial p$ (both in day$^{-1}$), with $\omega = \rho g w$, air density $\rho$, and gravitational acceleration $g$. Water detrainment (gm$^{-3}$day$^{-1}$) is likewise computed as $-\partial(\rho_w w)/\partial z$, using each convective cell's water density $\rho_w$ including vapour, cloud condensate, and precipitation.

We focus on the RCEMIP-I small-domain simulations (about 100 × 100 km, 1 km grid spacing, nominally 74 levels to 33 km), whose domains suppress convective self-aggregation and so isolate the processes central to the stability-iris hypothesis. We also briefly examine the large-domain simulations (6000 × 400 km, 3 km grid spacing), in which self-aggregation occurs. We use the models analyzed by Williams and Jeevanjee (2025), excluding MESONH due to a known humidity diagnostics issue (Mischell, 2026) and SAM_CRM, whose peak detrainment exceeds three times that of any other model and would dominate ensemble means. This leaves six models: CM1, DAM, SCALE, UCLA-CRM, UKMOi-vn11.0-CASIM and WRF_COL_CRM. For WRF_COL_CRM, which lacks archived clear-sky fluxes, clear-sky radiative cooling is taken as the all-sky heating of its cloud-free grid cells.

## 3. Results

### *3.1. Lack of a clear anvil amount response to warming across models*

We first examine how anvil cloud amount responds to warming across the assessed RCEMIP-I simulations. If the stability-iris mechanism represents a fundamental control on anvil amount, its predicted reduction should emerge consistently in models that capture relevant radiative-convective processes while minimizing complicating influences such as large-scale circulation. Yet Wing et al. (2020) and Stauffer and Wing (2022) found inconsistent anvil fraction responses across RCEMIP-I models. We here test whether this inconsistency holds across varied metrics of anvil amount, to establish whether a robust response exists before asking why one is absent.

Figure 1 presents the anvil amount responses of the small-domain RCEMIP-I simulations in the ways typically used to support the stability-iris hypothesis. In height coordinates ($f_c(z)$; Fig. 1a), the 305 K ensemble mean (red dashed line) peaks at a lower cloud fraction than the 295 K ensemble mean (black solid line), but about 3 km higher. To compare anvils at equivalent levels, we align each simulation by distance from its level of peak anvil amount ($f_c(z\text{-}z_{peak})$; Fig. 1b), defined as the cloud fraction maximum colder than 250 K. Because anvils rise nearly isothermally with warming, we also view responses along isotherms ($f_c(T)$; Fig. 1d; Bony et al., 2016; Jeevanjee, 2022), a framework we later use to link anvil responses to quantities with simple behavior along isotherms. In both frameworks, the cloud fraction response to warming is inconsistent across models (Fig. 1c,e): at most levels, only 2-3 of the six models show a reduction in anvil amount exceeding 1 % per kelvin of surface warming.

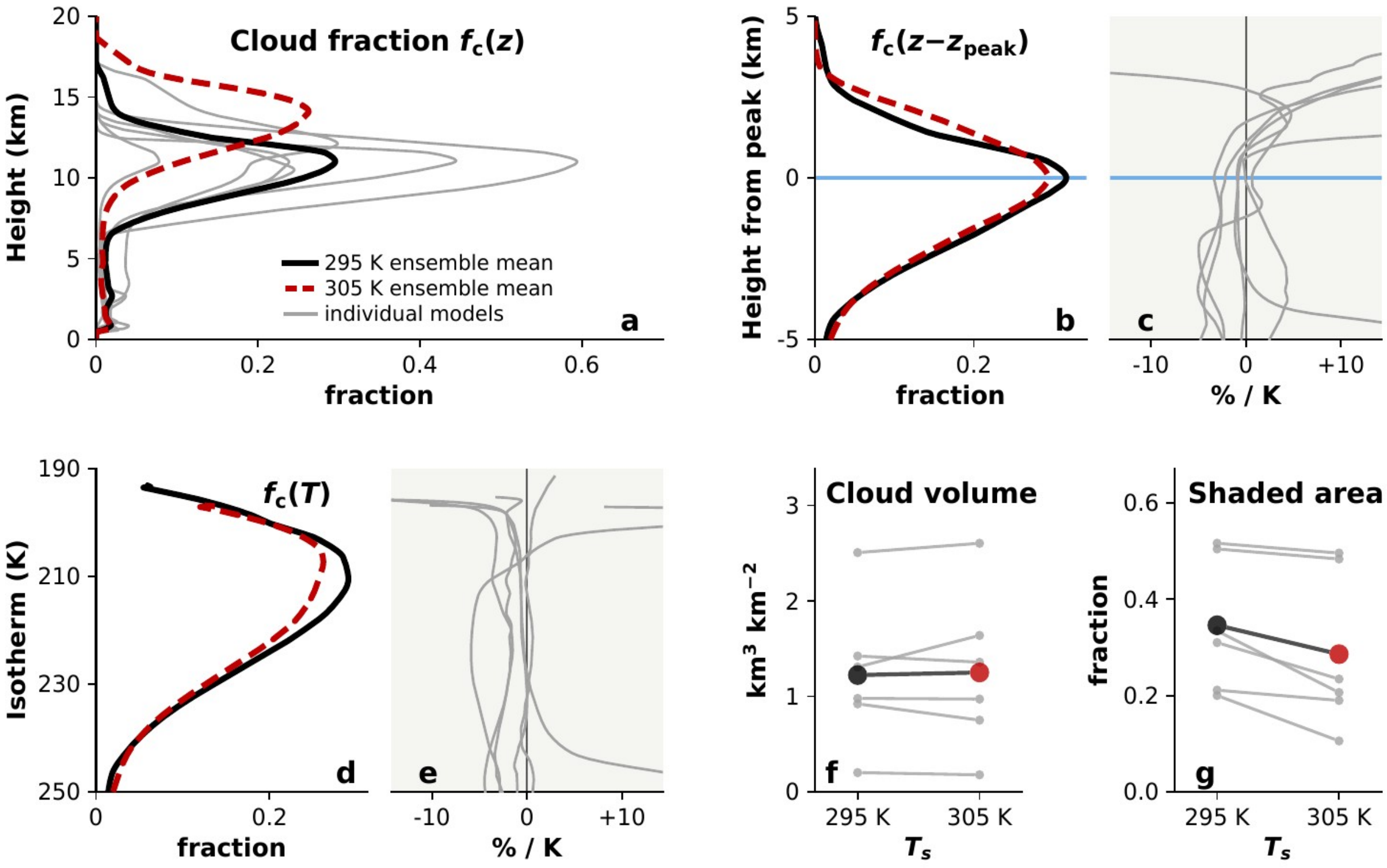

**Fig. 1** | Cloud fraction as provided in the RCEMIP-I archive for the six small-domain models, along altitudes (a), relative to the level of peak anvil amount (b), and along isotherms (d), with percent-per-kelvin sensitivities to surface warming (c,e). Cloud volume (f) and shaded area fraction (g) are integrals upward from 5 km below the peak level, and all

quantities are domain-mean. Black and red lines and markers show ensemble means at $T_s$ = 295 K and 305 K where at least five of the six models have data. Grey lines show individual models (at 295 K in a), grey dots individual simulations, and blue lines the level of peak anvil amount.

Anvil volume measures anvil amount through both its horizontal and vertical extent. Integrating cloud fraction upward from 5 km below the level of peak anvil amount, we find that the anvil volume per unit surface area ($km^3\ km^{-2}$; Fig. 1f) shows no consistent response across models, and its ensemble mean is nearly unchanged. Relative to the anvil peak, a slight reduction in cloud fraction near the peak is offset by a slight increase above it (Fig. 1b,c). The ensemble-mean reduction in peak anvil amount therefore appears to reflect mainly a vertical redistribution of anvil cloud rather than a loss.

The small-domain RCEMIP-I models therefore show no robust anvil amount reduction under surface warming, consistent with Wing et al. (2020) and Stauffer and Wing (2022). We note two caveats. First, the radiatively influential part of anvil cloud may decline even if total anvil amount does not, consistent with models producing optically thinner anvils with warming (Sokol et al., 2024). We test this by calculating shaded area as the fraction of the domain where condensate integrated upward from 5 km below peak anvil level exceeds 2 $g/m^2$, capturing most radiatively active anvil cloud (Gasparini et al., 2019, 2025; Deutloff et al., 2025). This fraction falls in all six models (Fig. 1g). Recalculating cloud fraction from the 3D snapshots with a uniform threshold across models gives a consistent result: little reduction for a lenient threshold ($10^{-8}\ g/m^3$), but slightly more than in Fig. 1 for a strict one ($10^{-2}\ g/m^3$; Fig. S1). Second, convective self-aggregation could also alter anvil amount. Large-domain RCEMIP-I simulations, which permit self-aggregation, show stronger but still inconsistent reductions (Fig. S2), whereas Bony et al. (2016) found self-aggregation was not responsible for their anvil reduction. The simulations therefore do not cleanly establish the decline that the stability-iris hypothesis proposes, motivating us to test whether its physical basis holds.

*3.2 Approximate invariance of convective motions along isotherms under warming*

Intuitively, the enhanced stability that tropospheric features such as anvil clouds experience as they rise with surface warming might be expected to weaken convectively driven motions. Because these motions drive the detrainment of air and water into the upper troposphere, such weakening would reduce anvil cloud amount. Yet this effect has not been rigorously established. In this section, we show that subsidence velocity along isotherms lacks a robust response to warming in the RCEMIP-I models and demonstrate theoretically that this is expected; Section 3.3 then uses these insights to infer how anvil-level detrainment responds to warming. To ease interpretation, we use geometric coordinates, and address the pressure coordinate formulation of Bony et al. (2016) in Section 3.4.

Viewing the RCEMIP-I output along isotherms reveals a key result (Fig. 2a,b): at most upper-tropospheric isotherms, subsidence velocity at 305 K is similar to that at 295 K. To show why approximate invariance is expected beyond these simulations, we first express the thermodynamic approximation for subsidence from Eqn. 1 in geometric coordinates, with $w$ here positive for subsiding air:

$$w = \frac{g Q_r}{T N^2} \qquad \text{Eqn. 3}$$

Here, $g$ is gravitational acceleration, $T$ is temperature, $N^2$ is height-coordinate static stability, and $Q_r$ is clear-sky radiative cooling as in Eqn. 1 (here positive for cooling). Noting that static stability $N^2 = \frac{g}{T}(\Gamma_d - \Gamma)$ , where $\Gamma$ is the atmospheric lapse rate of the environment and $\Gamma_d$ is its value in dry conditions, we rewrite Eqn. 3 following Jeevanjee (2022):

$$w = \frac{Q_r}{\Gamma_d - \Gamma} \qquad \text{Eqn. 4}$$

From here, the expression for $w$ simplifies most readily where the environmental lapse rate responds to warming like that of a moist adiabat ($\Gamma_m$), so that $\Gamma$ can be replaced by $\Gamma_m$ in Eqn. 4. This approximation is reasonable for isotherms warmer than about 225 K (Fig. S3), which contain the lower portion of anvils. A standard formula (Bohren and Albrecht, 1998) can then be rearranged to give

$$\Gamma_d - \Gamma_m = \frac{g}{c_p} \frac{\frac{L^2 r_s}{c_p R_v T^2} - \frac{L r_s}{R_d T}}{1 + \frac{L^2 r_s}{c_p R_v T^2}} \qquad \text{Eqn. 5}$$

where $r_s$ is water vapour mixing ratio at saturation, $L$ is the latent heating constant of condensate, $R_d$ is the gas constant for dry air, $R_v$ is the same for water vapour, and $c_p$ is the specific heat capacity of air at constant pressure. In the relatively dry upper troposphere the denominator $1 + L^2 r_s / (c_p R_v T^2)$ can be effectively approximated as unity. Simplifying accordingly, then factoring out the term $L r_s / (R_d T)$ gives

$$\Gamma_d - \Gamma_m \approx \frac{g L r_s}{R_d c_p T} \left( \frac{L R_d}{R_v c_p T} - 1 \right) \qquad \text{Eqn. 6}$$

The critical realization is that, along an isotherm, $r_s$ is the only term in Eqn. 6 that will change as an isotherm rises, and so the response of the actual $\Gamma_d$ - $\Gamma$ is likely to be largely proportional to that of $r_s$. Because $r_s = \rho_{v,s} / \rho$, where $\rho_{v,s}$ is vapour density at saturation, and the atmosphere's vapour density structure rises with surface warming to stay closely invariant along isotherms (Jeevanjee & Romps, 2018), this yields

$$\Gamma_d - \Gamma \propto \frac{1}{\rho} \qquad \text{Eqn. 7}$$

Both this relationship and our Eqn. 6 approximation are readily verified with the RCEMIP-I simulations (Fig. S3). Further, $Q_r$ measures cooling as a temperature tendency yet it is instead the volumetric cooling rate ($Q_V$, in W/m$^3$) that exhibits approximate isothermal invariance to surface warming (Jeevanjee, 2022). The conversion is a direct function of air density $\rho$, bringing us to our next expectation:

$$Q_V = \rho c_p Q_r \quad \Rightarrow \quad Q_r \propto \frac{1}{\rho} \qquad \text{Eqn.8}$$

Hence, along these isotherms both $Q_r$ and $\Gamma_d$ - $\Gamma$ will increase in response to surface warming, as the isotherms move into lower density air in which each unit of radiative and latent energy has an enhanced influence on temperature. Because $w$ is predicted by both $Q_r$ and $\Gamma_d$ - $\Gamma$ (Eqn.

4) and their main sensitivities directly cancel, $w$ along these isotherms is to first-order expected to be invariant to surface warming:

$$(w)_T \text{ nearly invariant} \qquad \text{Eqn. 9}$$

This thermodynamic reasoning cleanly applies at isotherms where moist adiabatic assumptions are predictive, which for this case span most of the free troposphere, at levels where $w$ sets the supply of air to anvils aloft but a minority of anvil volume is present. Notably, the simulations show the same invariance to hold aloft as well: no upper-tropospheric isotherm shows a consistent $w$ response across models (Fig. 2b). This is despite predictions from $Q_r/(\Gamma_d - \Gamma)$ failing at levels where anvils and convective overshoots complicate the heat budget: at isotherms colder than 225 K this framework predicts a reduction of $w$ that the resolved motions do not show (Fig. S4; relative to the anvil peak in Fig. S5), indicating that $w$ is maintained by cooling from processes other than clear-sky radiation. This may reflect cloud radiative heating weakening with warming (Fig. S6), consistent with anvils becoming optically thinner (Sokol et al., 2024), or convective overshoots transporting cold air more efficiently. Simple thermodynamic arguments therefore predict little change where they cleanly apply, and where they do not, the density effects on radiative cooling and lapse rate still cancel while additional factors also offset each other. In Section 3.3, we propose that water cycle constraints may explain why the isothermal invariance of convective motions holds this generally.

Viewing $w$ relative to the level of peak anvil amount (Fig. 2g,h), there is similarly no clear response where anvil amount peaks or below. However, from 1 km above the peak, models show $w$ to strengthen. Here the lapse rate decreases with warming (Fig. S7), so levels at fixed distances above the peak shift onto warmer isotherms (Fig. S8) where subsidence is stronger (Fig. 2a). Convectively driven motion is thus anchored to temperature rather than height, and warming relocates these motions upward without a clear reduction. As $w$ also equals the integrated upward motion of convective plumes, this invariance appears to contradict the widely reported weakening of convection, which we reconcile in a separate study (McGraw & Gasparini, 2026). Whether detrainment at anvil levels remains similarly unchanged is examined next.

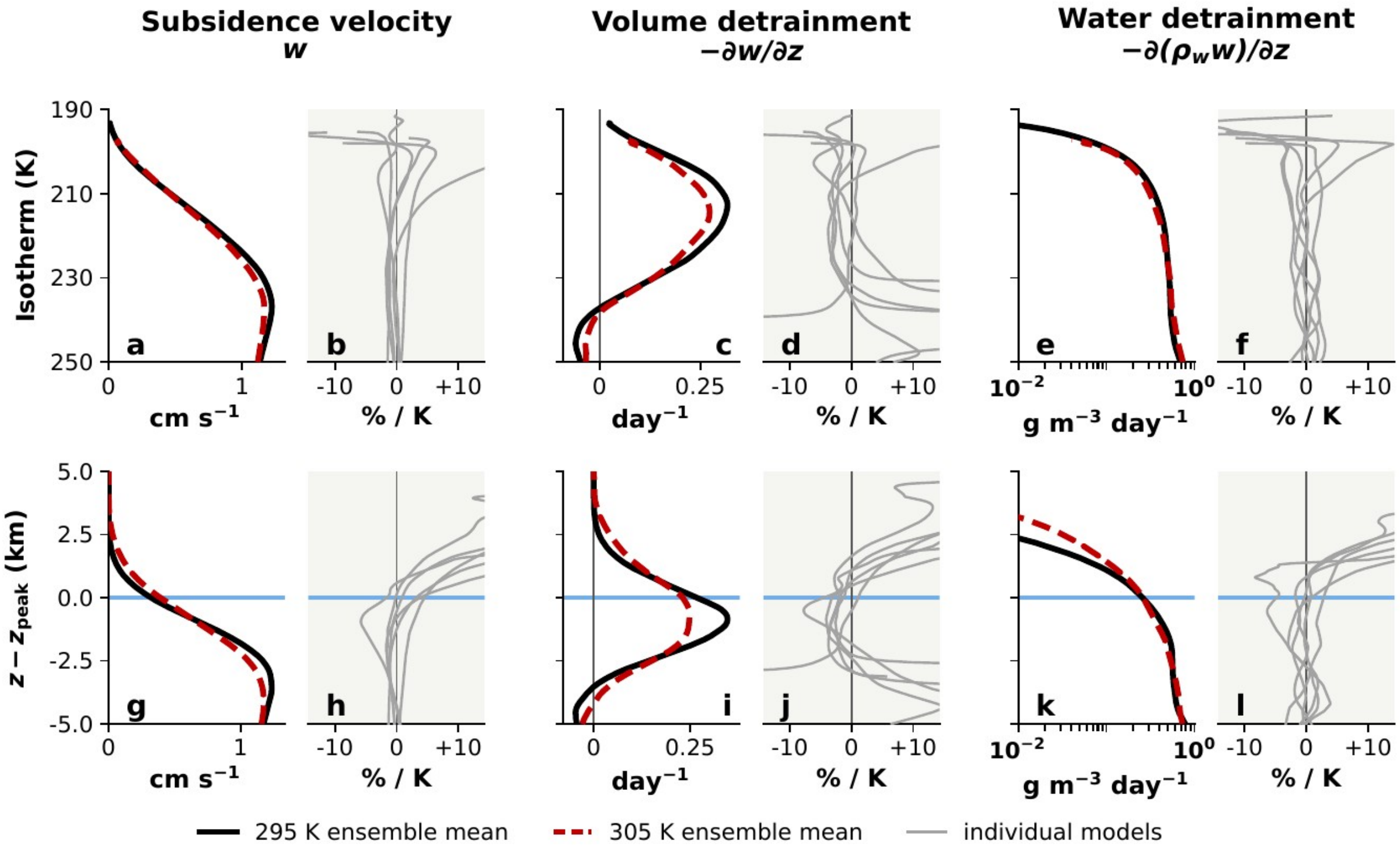

**Fig. 2** | Subsidence velocity *w* (left column), volume detrainment -∂*w*/∂*z* (middle column) and water detrainment $-\partial(\rho_w w)/\partial z$ (right column) along isotherms (top row) and relative to the level of peak anvil amount (bottom row) across RCEMIP-I, alongside their sensitivities to surface warming. All quantities are calculated directly from the resolved convective motions. Water detrainment includes water vapour and all condensate, and is shown on a logarithmic axis. Detrainment sensitivities are shown only above the base of the outflow layer, omitting negative values. Plot elements are as described in the Fig. 1 caption.

*3.3 Why does warming not produce a robust reduction in air detrained at anvil level?*

Anvils are fed by air that convective plumes detrain as their upward volume flux decreases with height, which we measure as the volume detrainment -∂*w*/∂*z*. This quantity closely matches the pressure-coordinate detrainment ∂*ω*/∂*p* of Bony et al. (2016), which gives the net volume of air leaving convection per unit volume and time, or equivalently the fraction of a layer's horizontal area that newly detrained air spreads over per unit time. The two differ only by a density term reflecting expansion of rising air that is small in the upper troposphere (Fig. S9). Detrainment is tied directly to the subsidence *w* assessed in Section 3.2: at each isotherm, *w* equals the total volume of air detrained above it, after this same expansion adjustment (Fig.

S10). Moreover, since $-\partial w/\partial z = \Gamma\ \partial w/\partial T$, detrainment along isotherms or at any height relative to the anvil peak follows from the *w(T)* profile, the lapse rate, and the location of the anvil peak alone. The approximate invariance of *w* along isotherms therefore provides an initial expectation that anvil-level detrainment changes little with warming.

In the simulations, the volume detrainment $-\partial w/\partial z$ indeed changes modestly along isotherms, with an ensemble-mean reduction near 1.5 %/K within 210-220 K, reflecting reductions in most but not all models (Fig. 2c,d). Following the relation above, this reduction has two sources. First, the lapse rate decreases with warming (Fig. S7), so each Kelvin spans a deeper layer. This geometric effect spreads the same detrained volume over more height, lowering detrainment per unit volume without reducing the total air detrained between isotherms. Second, the modest deviations of *w* from invariance on average reduce how much *w* changes per kelvin across the anvil layer (Fig. 2a).

On average the anvil peak warms by about 0.1 K per kelvin of surface warming, though inconsistently across models (Fig. S8; see also Wing et al., 2020). Could this slight warming of anvils, as expected from PHAT (Zelinka & Hartmann, 2010), substantially reduce detrainment at anvil level? There is no clear reason to expect so, since that would require the anvil peak to shift in temperature relative to the detrainment profile, whereas PHAT rests on the two being linked. Consistent with this, detrainment at and shortly below the anvil peak declines about as much as along isotherms (Fig. 2i,j). Above the peak, by contrast, levels shift onto increasingly warmer isotherms with height, as subsidence does (Section 3.2), and detrainment hence increases in all models. At heights above the anvil peak, increasing stability thus coincides with *stronger* detrainment, the opposite of what the stability-iris hypothesis anticipates.

The clear-sky radiative estimate $-\partial(Q_r/(\Gamma_d - \Gamma))/\partial z$, the geometric counterpart of $D_r$ (Eqn. 2), cannot capture the limited detrainment response: it responds far more strongly to warming than detrainment quantified from the resolved motions, both along isotherms (Fig. S4c,d) and at the anvil peak (Fig. S5c,d). This reflects inaccuracies of $Q_r/(\Gamma_d - \Gamma)$ for predicting *w* discussed in Section 3.2. We propose that the water cycle may constrain air detrainment more tightly than clear-sky thermodynamics, and be essential to maintaining the approximate invariance of both $w$ and $-\partial w/\partial z$. In one of the most robust expectations for the warming tropical troposphere, water density stays invariant along isotherms, as the simulations show (Fig. S12). As this density balances detrained water against its sinks, it implies invariant water detrainment insofar as the sinks, in sum, change little. Simulated water detrainment indeed stays close to isothermal invariance, and more so than air detrainment (Fig. 2e,f; split by phase in Fig. S11). The approximate invariance of detrained air may thus reflect that of detrained water, consistent with water vapour radiative cooling shaping the buoyancy gradients that drive convection (Hartmann & Larson, 2002; Jeevanjee & Romps, 2018). Conversely, the modest ensemble-mean decline in detrained air, if meaningful, implies each unit volume carries slightly more water with warming, possibly reflecting intensified convective updrafts (Singh & O'Gorman, 2016; Jenney et al., 2020).

The response of anvil amount near its peak (Fig. 1b,c) resembles that of air detrainment (Fig. 2i,j), declining at and below the peak and increasing aloft, but is about half as large on average. Other factors also shape the anvil amount response, including the more invariant detrainment of water, and potentially changes in the sinks of anvil cloud (Beydoun et al., 2021; Sokol & Hartmann, 2022). Either way, the inconsistent response of anvil amount reflects a convective supply of air and water that lacks a clear reason to decline substantially, rather than the robust stability-driven reduction proposed by the stability-iris hypothesis.

### *3.4. Breakdown of the stability-iris hypothesis*

We now reconcile the approximate invariance of convectively driven motion and detrainment along isotherms and at peak anvil level (Sections 3.2-3.3) with the expectations of the Bony et al. (2016) stability-iris hypothesis, which is formulated in pressure coordinates. We first represent the pressure-coordinate static stability used by Bony et al. (2016) as follows (Holton, 2004):

$$S = \frac{\Gamma_d - \Gamma}{\rho g} \qquad \text{Eqn. 10}$$

The numerator responds to surface warming in inverse proportion to ρ (Eqn. 7), so together with the explicit 1/ρ dependence from the denominator, $S \propto 1/\rho^2$. By Eqns. 1 and 8, the response of ω then follows that of $Q_r$ (∝ $1/\rho$) divided by that of S, yielding the expectation that $\omega \propto \rho$. Finally, along an isotherm the ideal gas law gives $\rho \propto p$, so the $\rho$ dependence of the estimated $\omega$ is approximately canceled by the $p$ dependence introduced when taking its pressure derivative. Hence, where the predictive framework is useful, $\partial\omega/\partial p$ along an isotherm is expected to change little with surface warming:

$$S \propto \frac{1}{\rho^2} \quad ; \quad \omega \propto \rho \quad ; \quad \left(\frac{\partial \omega}{\partial p}\right)_T \text{ nearly invariant} \qquad \text{Eqn. 11}$$

Eqn. 11 poses a fundamental challenge to the stability-iris hypothesis: *the increase in static stability as anvils rise does not systematically alter the detrainment of air into anvil levels*, breaking the proposed control on anvil amount.

The simulations support Eqn. 11 (Fig. 3). Air density along isotherms decreases by about 3 %/K in all models, implying roughly +6 %/K for $S$ and -3 %/K for $\omega$ (gold lines), which captures most of the simulated responses of both (Fig. 3a-d). The decline in $\omega$ is thus a density effect rather than a weakening of convectively driven motion, which remains nearly invariant along isotherms (Section 3.2). This density effect explains the theorized anvil reduction of 1-4 %/K of McKim et al. (2024), who took anvil amount as proportional to $\omega$ rather than $\partial\omega/\partial p$. By contrast, $\partial\omega/\partial p$ lacks a robust response along isotherms (Fig. 3e,f), as Eqn. 11 predicts, and closely mirrors the volume detrainment $-\partial w/\partial z$ (Fig. S9), including its increase above the anvil peak (Fig. 3k,l; Section 3.3).

The $Q_r/S$ predictor of Bony et al. (2016; our Eqn. 1) likewise follows the expected dependence on $\rho$ (Fig. S13). Like its geometric counterpart (Sections 3.2 and 3.3), however, it exaggerates the response of $\omega$ in the upper troposphere, and its derivative $D_r$ (Eqn. 2) departs strongly from simulated detrainment, with exaggerated declines at the coldest isotherms (compare Figs. 3 and S13). The decline in $D_r$ central to the stability-iris hypothesis hence likely partly reflects the limits of the clear-sky estimate in the upper troposphere, rather than a reduction in the detrainment that supplies anvils.

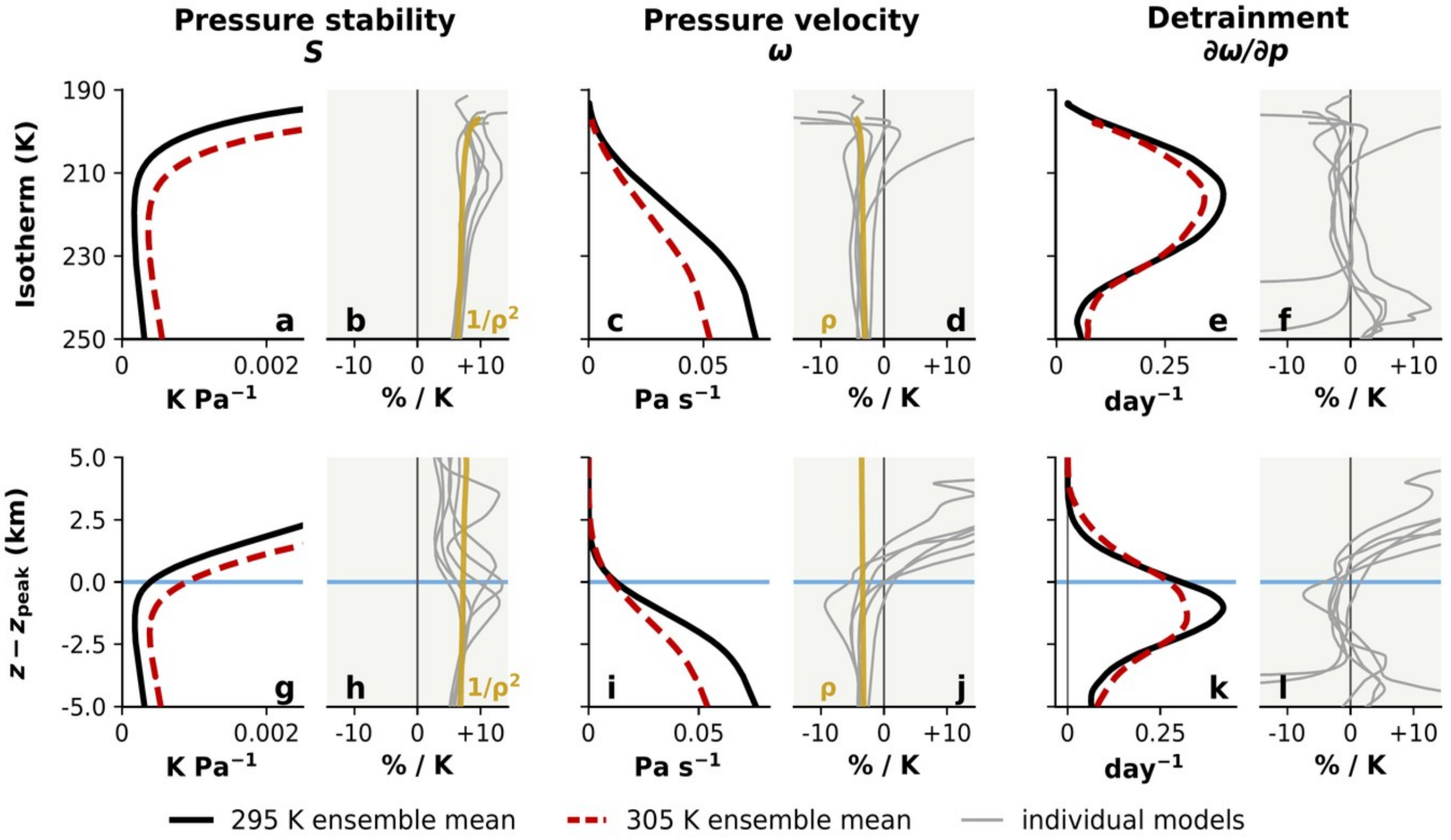

**Fig. 3 |** Key quantities of the stability-iris hypothesis across RCEMIP-I: pressure-coordinate static stability S (left column), the pressure velocity of the subsidence $\omega$ (middle column) and its associated net detrainment $\partial\omega/\partial p$ (right column), alongside their sensitivities to surface warming. $\omega$ and $\partial\omega/\partial p$ are calculated directly from the resolved convective motions, not from radiative cooling and stability (cf. Fig. S13). Gold lines show the density influence on $S$ and $\omega$ (Eqn. 11) using the ensemble-mean sensitivity of $\rho$. Plot elements are as described in the Fig. 1 caption.

The stability-iris mechanism therefore lacks a clear physical basis for anvil reduction, consistent with the absence of a robust anvil amount response (Section 3.1). The stability increase that underpins the hypothesis has motivated the expectation of an anvil control that does not exist: it arises largely from the decreasing density of air at a given isotherm, an influence that does not carry through to reducing the detrainment into anvil levels.

## 4. Conclusions

These findings reconcile the lack of robust support for a stability-driven reduction in anvil cloud amount: *the physics underlying the proposed mechanism instead point to little change* *in the supply of air to anvils*. This explains why, in the RCEMIP-I models, air and water detrainment at anvil levels change far less than the robust stability increase, and is consistent with their lack of a robust anvil reduction and thermodynamic expectations.

Observed decreases in anvil amount (Lindzen et al., 2001; Choi et al., 2017) remain debated (Hartmann & Michelsen, 2002; Williams & Pierrehumbert, 2017; Raghuraman et al., 2024). Recent studies link anvil changes to upper-tropospheric stability in present-day variability (Saint-Lu et al., 2020, 2022; Ito & Masunaga, 2022; Wilson Kemsley et al., 2024) and after volcanic eruptions (Saint-Lu et al., 2022). Crucially, associated stabilizations are fundamentally different from the density-driven stabilization experienced by rising anvils: interannual variability is dominated by El Niño circulation impacts, while volcanic aerosols reduce anvils by heating the upper troposphere and decreasing its depth (McGraw & Polvani, 2026). Our results show that a stability increase need not reduce the supply of air to anvils, calling for more careful causal interpretation of stability-anvil relationships in observations.

Our results shift attention from stability toward more complex controls on anvil formation and persistence, particularly ice microphysics and convective aggregation. These processes may be key to explaining the divergent simulated responses of anvil clouds to warming but are not fully represented in the idealized RCEMIP-I simulations considered here. Investigating more comprehensive simulations is therefore a crucial next step, and is already underway (e.g., Ohno & Satoh, 2018; Deutloff et al., 2026).

**Acknowledgments**

We thank the RCEMIP community for making their simulation output publicly available.

**Open access statement**

Data used in this study are available from the Radiative-Convective Equilibrium Model Intercomparison Project Phase 1 (RCEMIP-I) archive (Wing et al., 2018b). The RCEMIP-I simulation output analyzed here is publicly available through the RCEMIP data repository.

**Conflict of Interest**

The authors attest that they have no conflicts of interest.

## Supporting Information

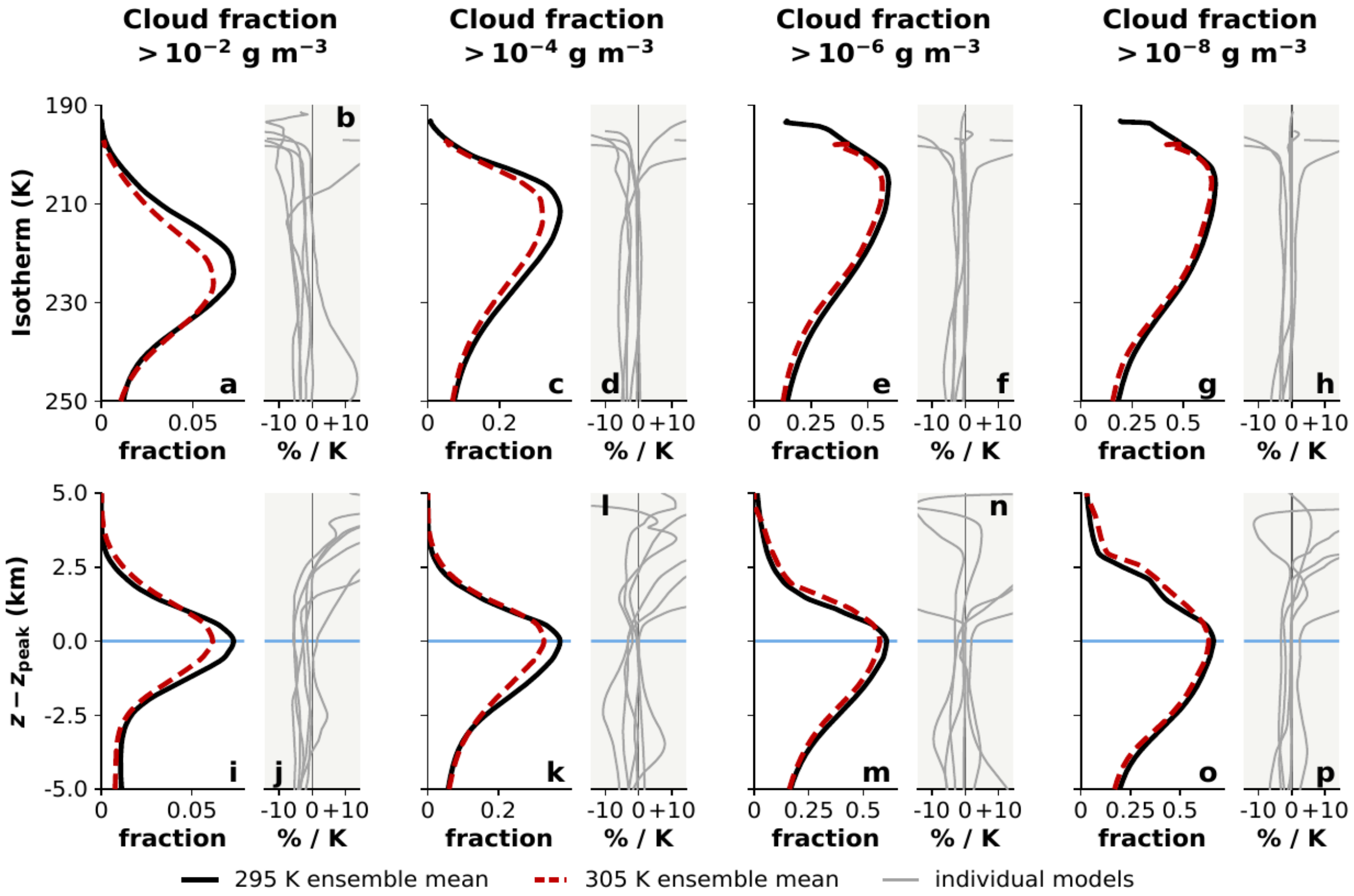

**Fig. S1** | As in Fig. 1b,c,d,e, but with cloud fraction recalculated from the archived 3D snapshots instead of taken as provided in the RCEMIP-I archive. Cloud fraction is recalculated for condensate (liquid+ice) density thresholds from $10^{-2}$ to $10^{-8}$ g $m^{-3}$ (columns), and in the bottom row the level of peak anvil amount is that of each threshold's own cloud fraction.

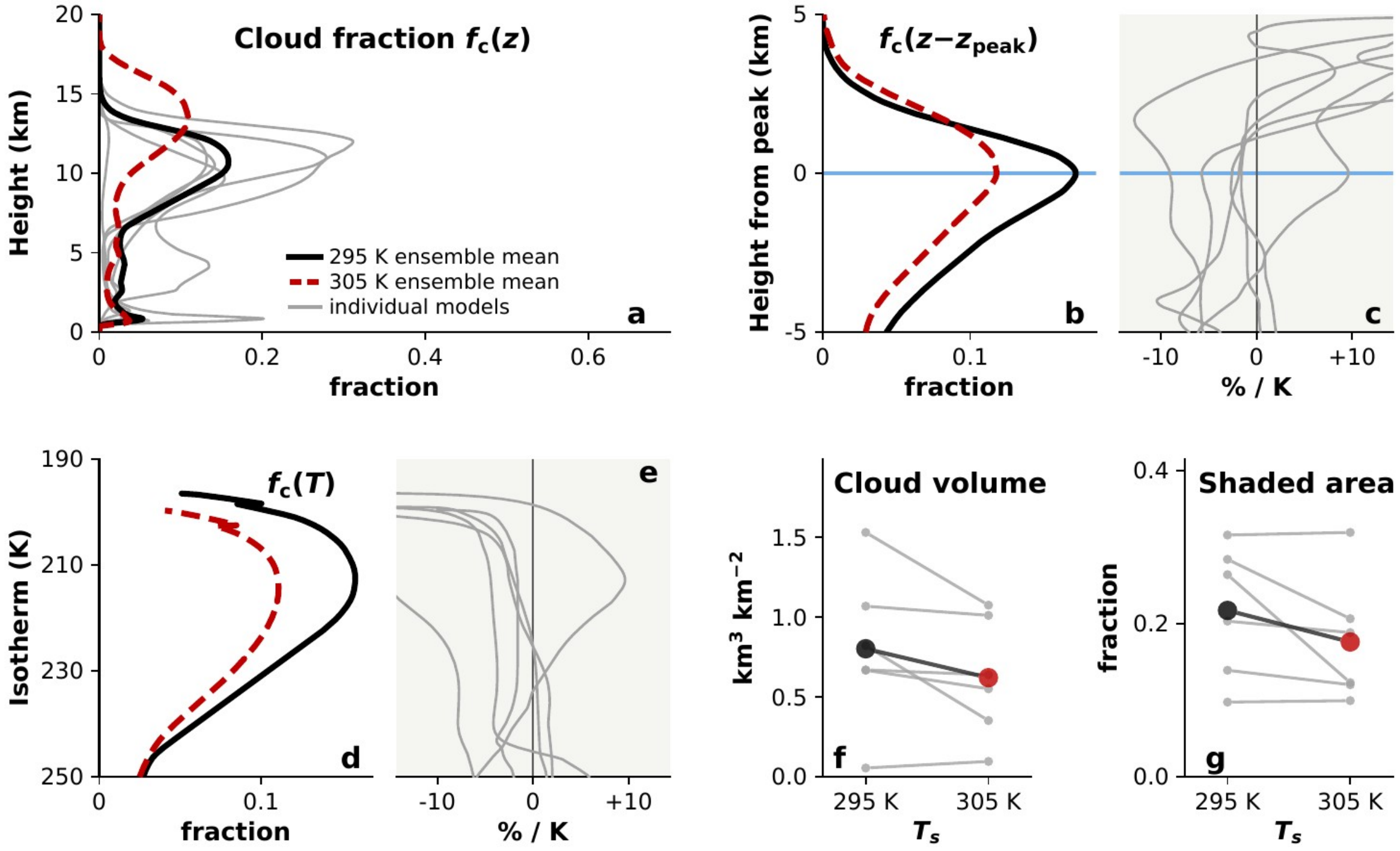

**Fig. S2** | As in Fig. 1, but using the large-domain RCEMIP-I simulations, which permit convective self-aggregation, rather than the small-domain simulations shown everywhere else in this study. The same six models are assessed in both ensembles.

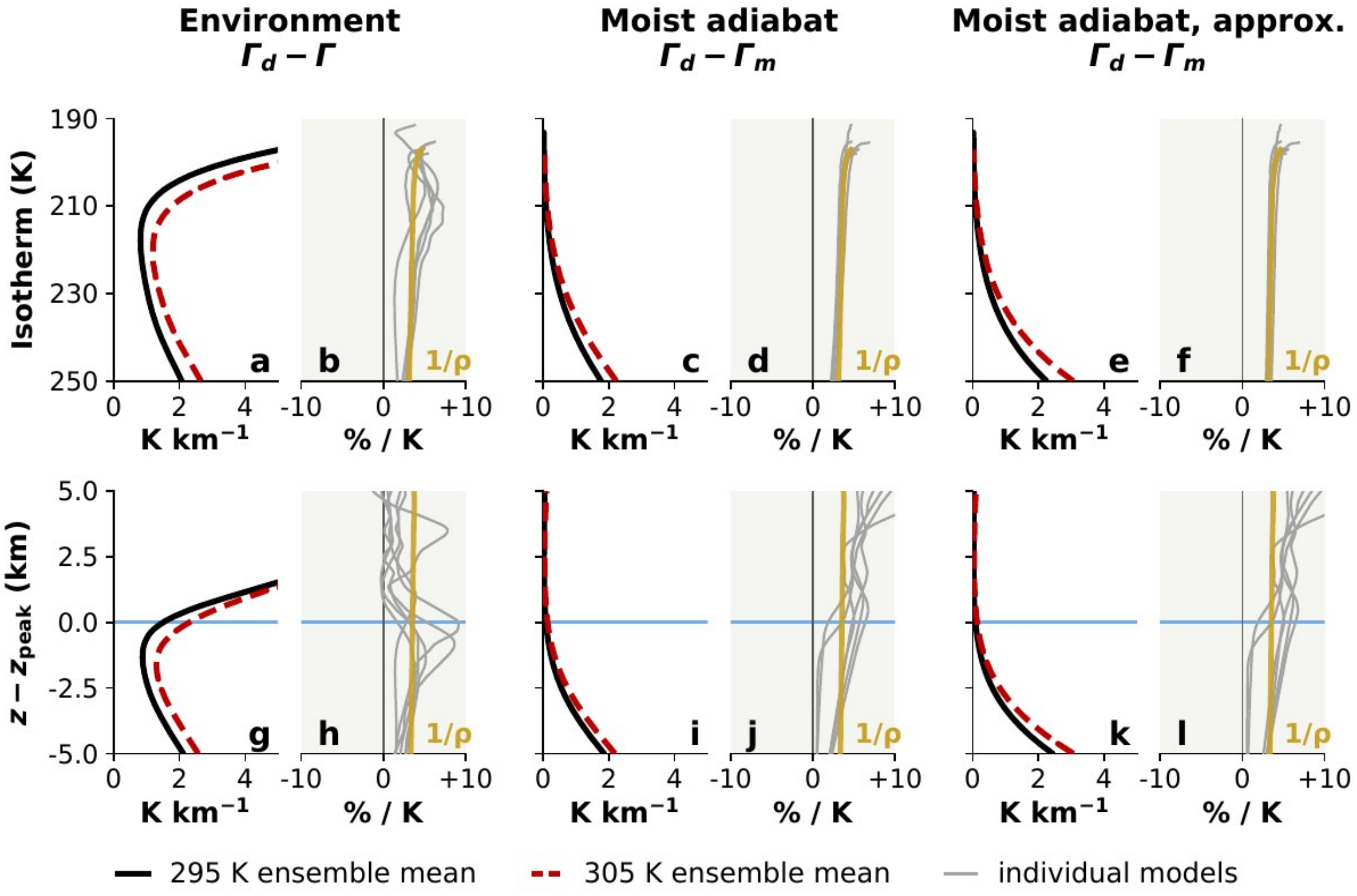

**Fig. S3** | The dry adiabatic lapse rate minus the simulated lapse rate ($\Gamma_d$ - $\Gamma$, the denominator of Eqn. 4; left column), minus the local moist adiabatic lapse rate from simulated temperature and pressure ($\Gamma_d$ - $\Gamma_m$, Eqn. 5; middle column), and minus our approximation of it (Eqn. 6; right column). Each is shown along isotherms (top row) and relative to the level of peak anvil amount (bottom row), with its sensitivity to warming. All are domain means. Gold lines show the ensemble-mean sensitivity of $1/\rho$, which largely sets each sensitivity and, along isotherms, fully sets that of the approximation. Other plot elements are as in Fig. 1.

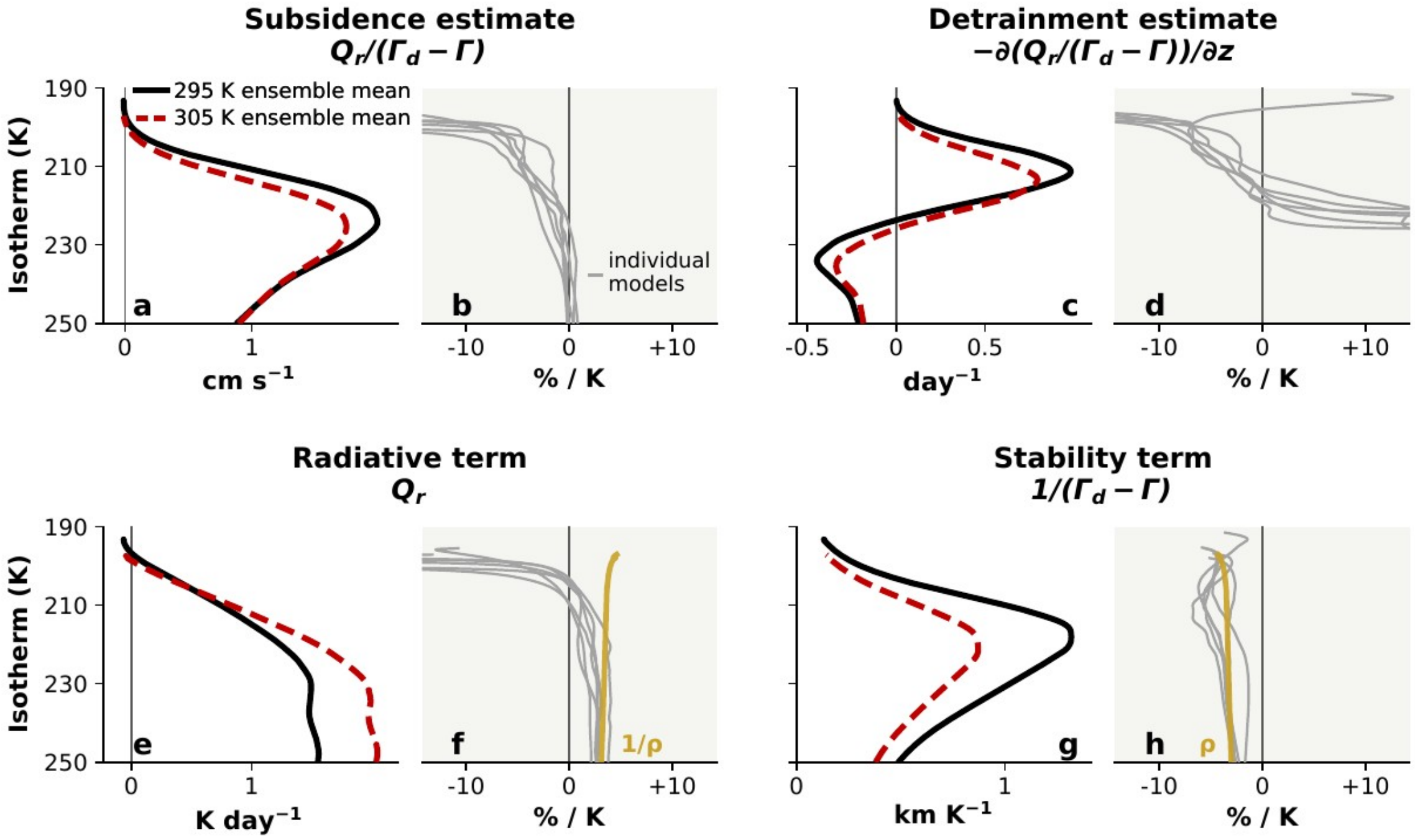

**Fig. S4** | Subsidence and detrainment estimated from domain-mean clear-sky radiative cooling and stability (Eqn. 4), along isotherms, for comparison with the directly calculated values in Fig. 2. Shown are the estimated subsidence $Q_r/(\Gamma_d - \Gamma)$ (a,b) and detrainment $-\partial(Q_r/(\Gamma_d - \Gamma))/\partial z$ (c,d), and the radiative term $Q_r$ (e,f) and stability term $1/(\Gamma_d - \Gamma)$ (g,h) from which they are built. For WRF_COL_CRM, which lacks clear-sky output, $Q_r$ is the all-sky radiative cooling of cloud-free grid cells. Gold lines show the influence of air density on $Q_r$ ($\propto 1/\rho$) and on $1/(\Gamma_d - \Gamma)$ ($\propto \rho$), which cancel in the estimated subsidence. Other plot elements are as described in the Fig. 1 caption.

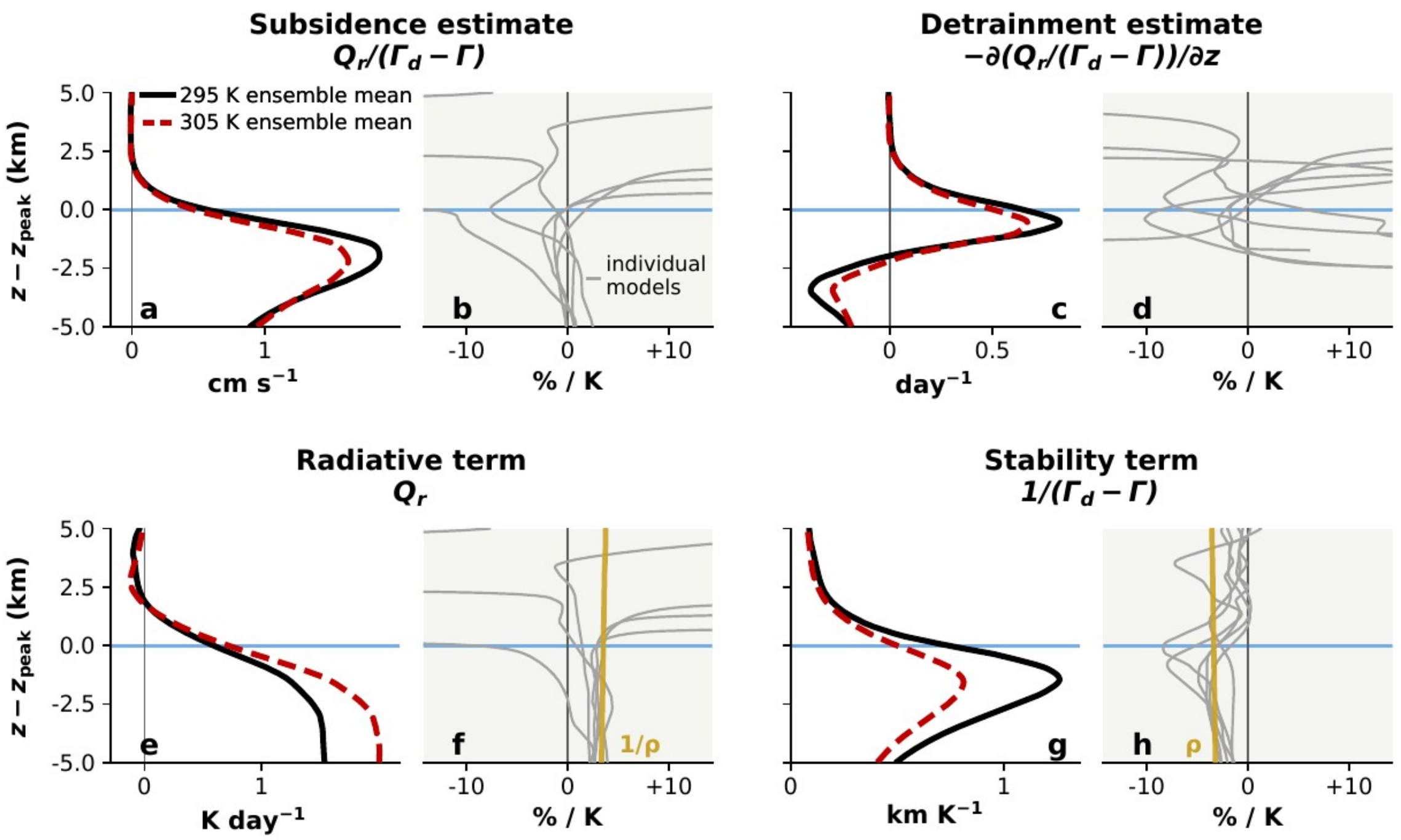

**Fig. S5** | As in Fig. S4, but relative to the level of peak anvil amount rather than along isotherms.

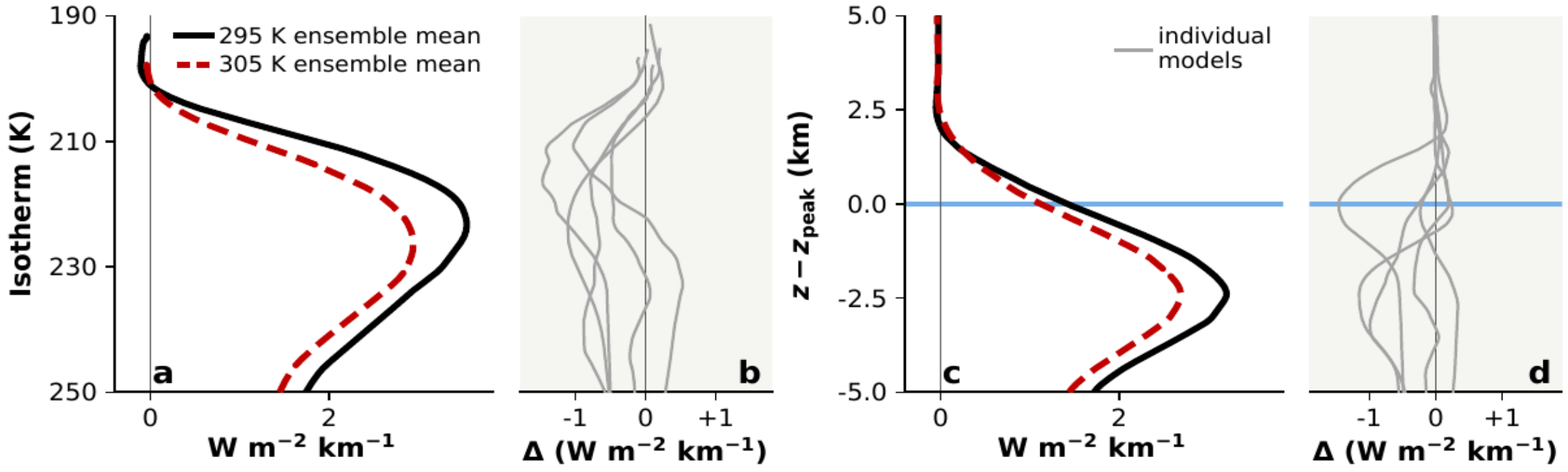

**Fig. S6** | Domain-mean cloud radiative heating per unit volume (all-sky minus clear-sky) along isotherms (a) and relative to the level of peak anvil amount (c), and its absolute change from 295 K to 305 K in each model (b,d). WRF_COL_CRM is excluded because it lacks clear-sky output. Plot elements are as described in the Fig. 1 caption.

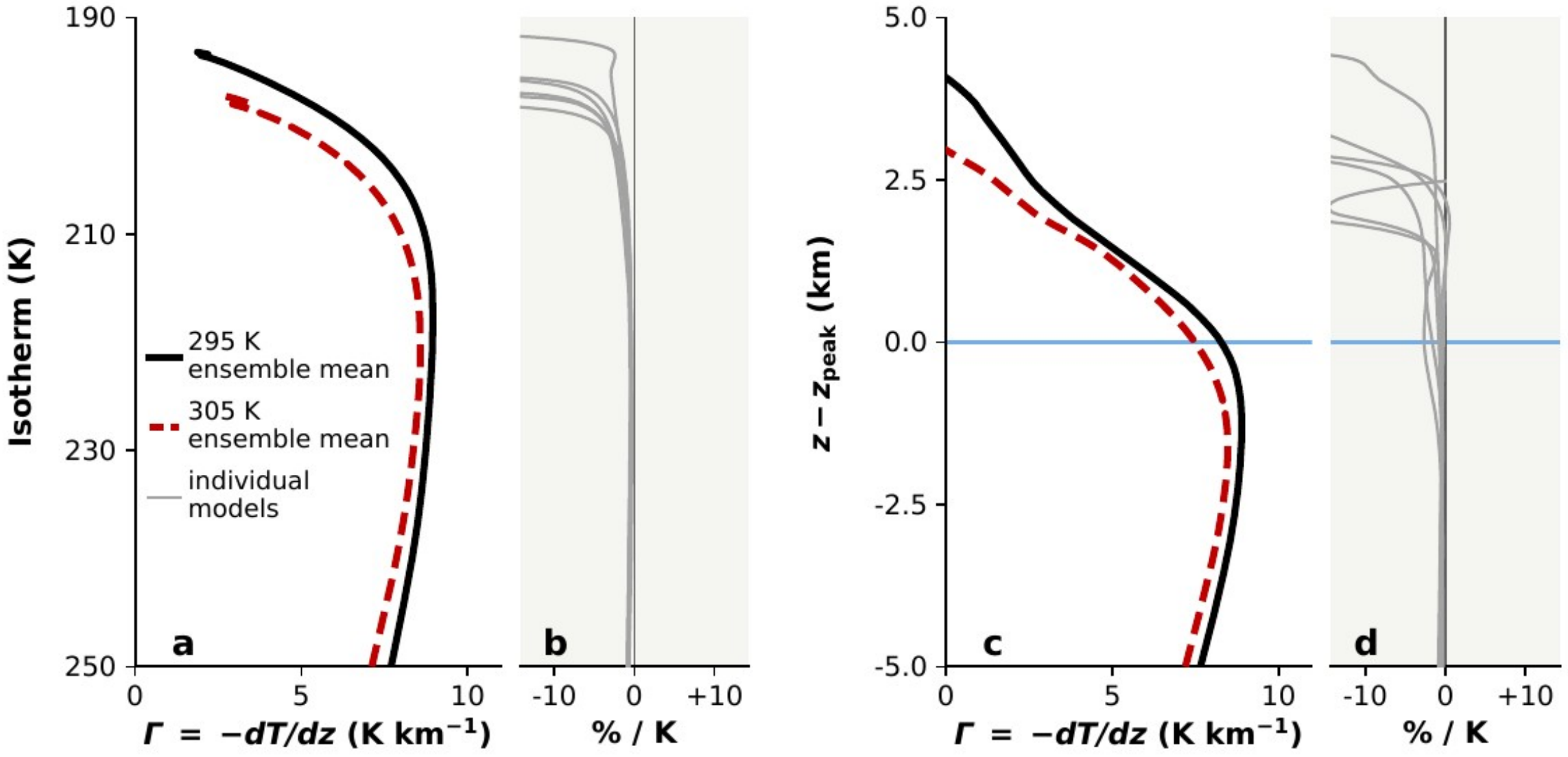

**Fig. S7** | The domain-mean lapse rate *Γ* = *-dT/dz* along isotherms (a) and relative to the level of peak anvil amount (c), alongside its %/K sensitivity to warming in each model (b,d). Plot elements are as described in Fig. 1.

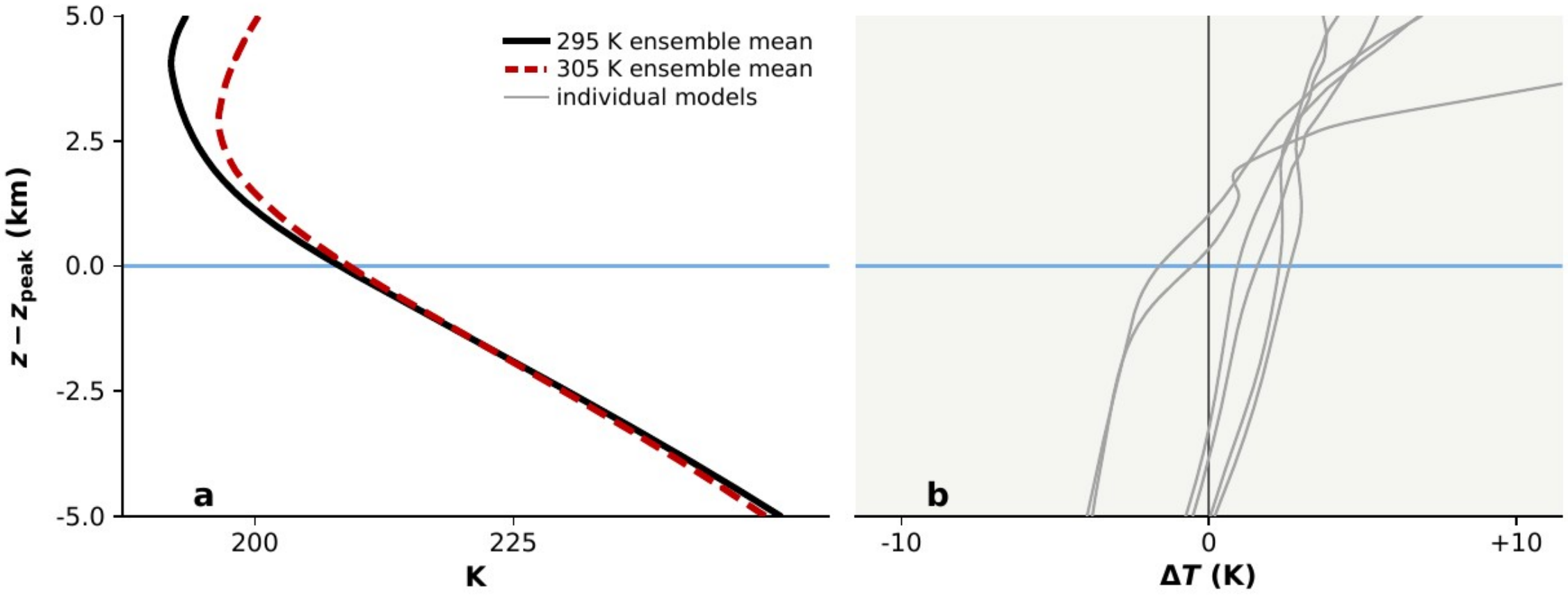

**Fig. S8 |** Domain-mean temperature relative to the level of peak anvil amount (a) and its change from 295 K to 305 K in each model (b). Panel (b) shows how far the anvil peak and the levels around it depart from an isothermal rise as the surface warms. Plot elements are as described in Fig. 1.

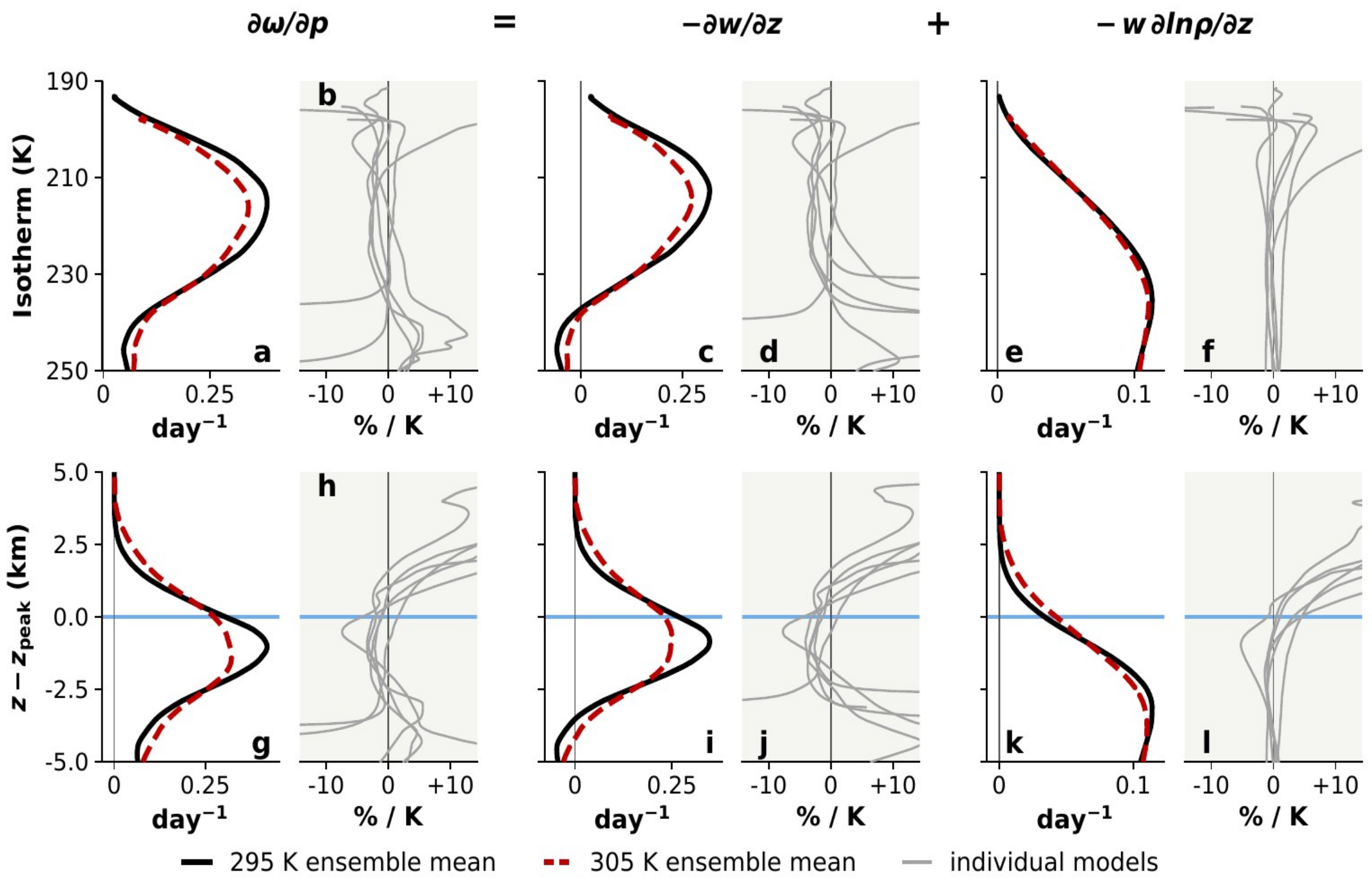

**Fig. S9** | Pressure-coordinate detrainment *∂ω/∂p* (left column; as in Fig. 3) equals the geometric-coordinate detrainment *-∂w/∂z* (middle column; as in Fig. 2) plus a density term reflecting the expansion of rising air, *-w∂lnρ/∂z* (right column). These values are calculated directly from the resolved convective motions. This shows that the two measures of detrainment, and their sensitivities to warming, are nearly the same in the upper troposphere. Plot elements are as described in Fig. 1.

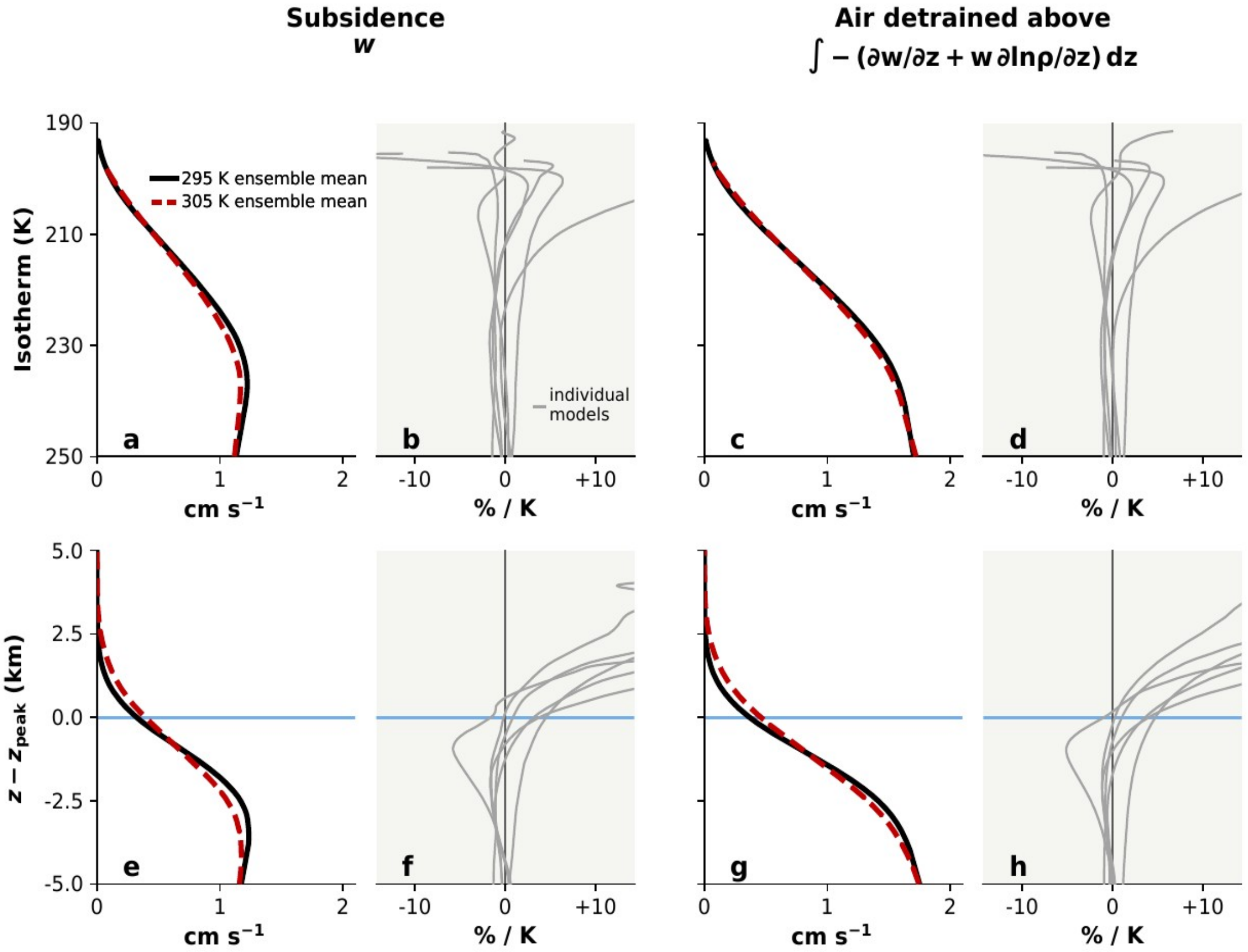

**Fig. S10** | Subsidence velocity *w* (left column; as in Fig. 2) and the volume of air detrained above each level (right column), ∫ -($\partial w/\partial z$ + $w\partial ln\rho/\partial z$) dz from that level to the model top, which is the pressure-coordinate detrainment $\partial\omega/\partial p$ (Fig. S9) integrated in height. The two columns differ only by the integrated density term of Fig. S9, reflecting the expansion of air as it rises. Both are calculated directly from the resolved convective motions. Plot elements are as described in Fig. 1.

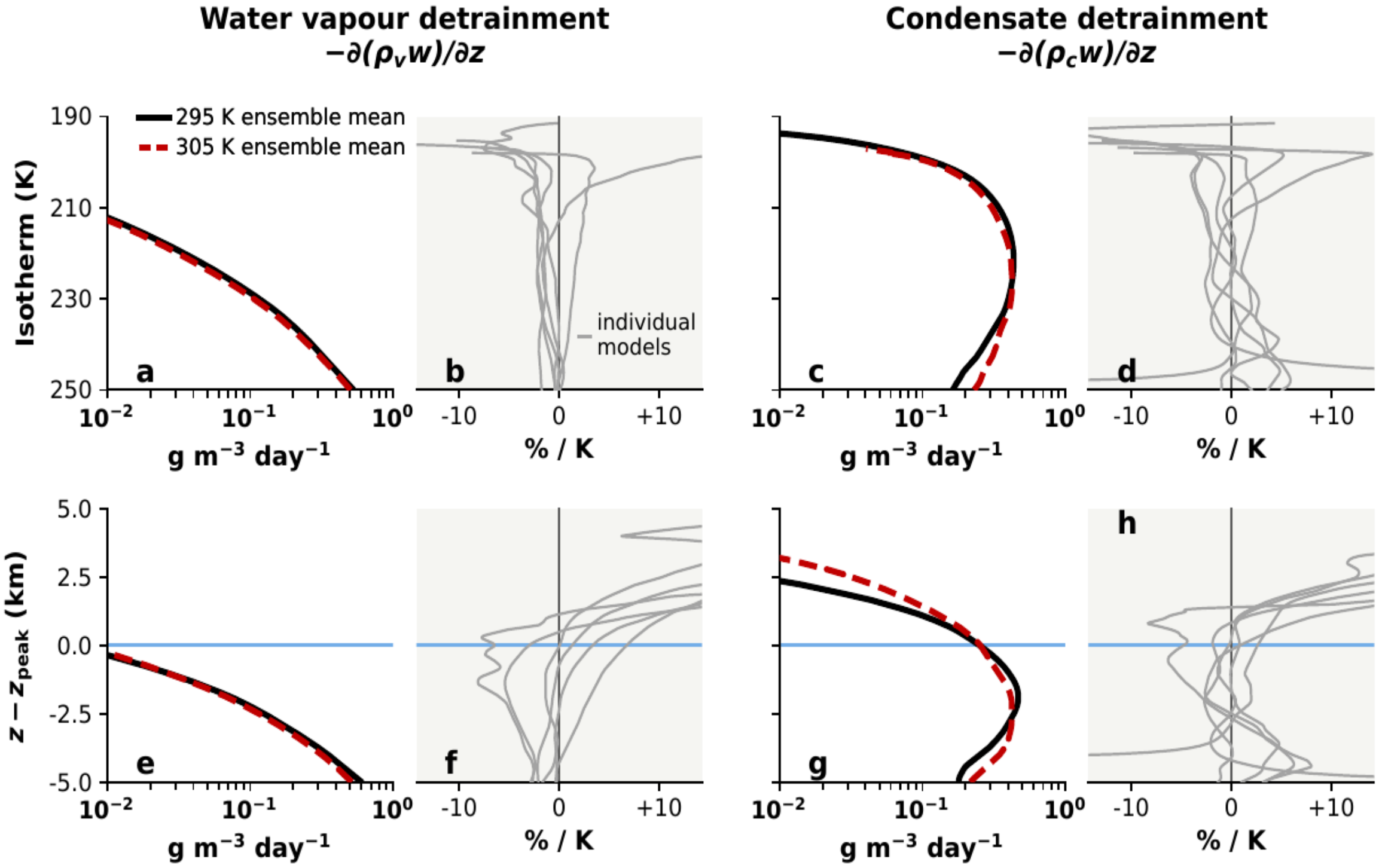

**Fig. S11** | As in the right column of Fig. 2, but with water detrainment shown separately for water vapour (left column) and condensate (right column, including both in-cloud and precipitating species).

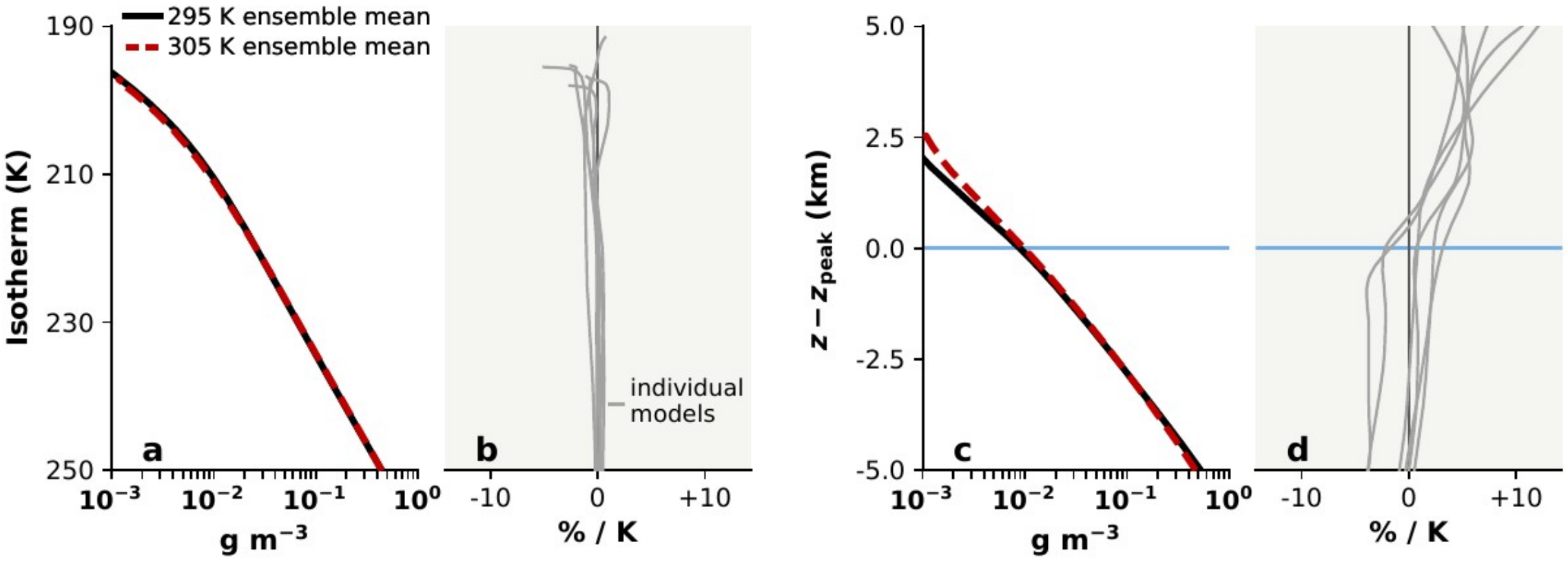

**Fig. S12** | Water density $\rho_w$ along isotherms (a) and relative to the level of peak anvil amount (c) across RCEMIP-I, alongside their sensitivities to surface warming (b,d). Water density is calculated as the domain-mean density of water vapour and all condensate (both in-cloud and precipitating). Plot elements are as described in the Fig. 1 caption.

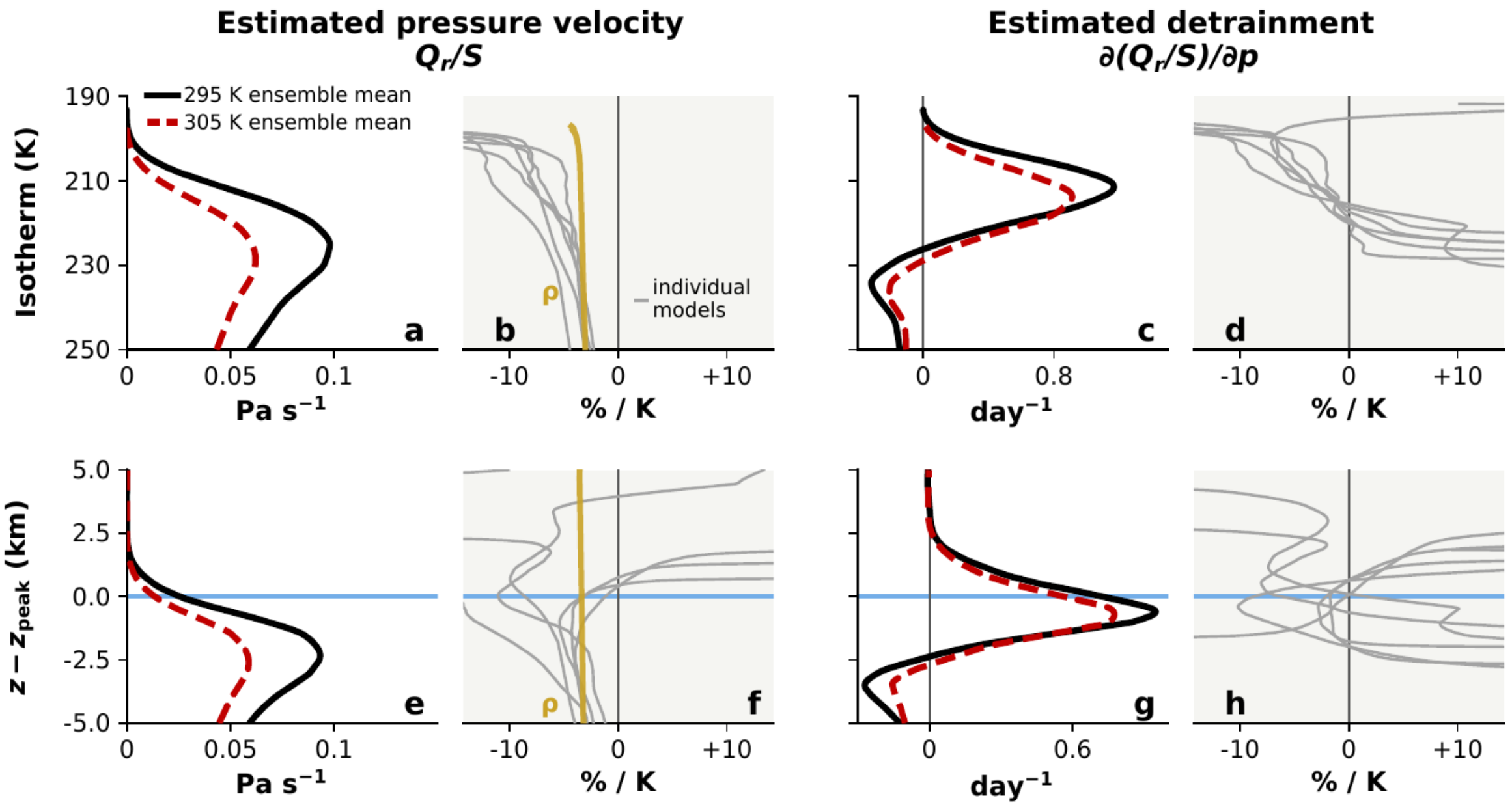

**Fig. S13** | The Bony et al. (2016) diagnostics, estimated from domain-mean clear-sky radiative cooling and static stability: the pressure velocity estimate $Q_r/S$ (Eqn. 1; left column) and its associated net detrainment $\partial(Q_r/S)/\partial p$ (Eqn. 2; right column), for comparison with the directly calculated values in Fig. 3. $Q_r$ is shown separately in Fig. S4. For WRF_COL_CRM, which lacks clear-sky output, $Q_r$ is the all-sky radiative cooling of cloud-free grid cells. Gold lines show the influence of air density on the estimated pressure velocity ($\propto \rho$). Other plot elements are as described in the Fig. 1 caption.